\documentclass[preprint,showpacs,showkeys,aps,prd,amsfonts,eqsecnum,nofootinbib,superscriptaddress]
{revtex4} 
\usepackage{graphicx,epsfig,amssymb}
\usepackage{axodraw}

\def\ptslash{{p_{t}}{\kern -1.5ex\hbox{/}}}
\def\gsim{\lower0.5ex\hbox{$\:\buildrel >\over\sim\:$}}
\def\lsim{\lower0.5ex\hbox{$\:\buildrel <\over\sim\:$}}

\newcommand{\re}{\mathop{\mathrm{Re}}}
\newcommand{\im}{\mathop{\mathrm{Im}}}
\newcommand{\Rslash}{R\kern -6.4pt\big{/}}

\begin{document}
%
\preprint{MADPH-05-1425,CUMQ/HEP-134,WM-05-108}
%
%
\title{Width Effects on Near Threshold Decays of the Top Quark 
\boldmath{$ t \to c WW,~cZZ$}  and of Neutral Higgs Bosons}
\author{Shaouly Bar-Shalom}
\email[]{shaouly@physics.technion.ac.il}
\affiliation{Department of Physics, Technion-Israel Institute of Technology, 32000 Haifa, Israel}
\author{Gad Eilam}
\email[]{eilam@physics.technion.ac.il}
\affiliation{Department of Physics, Technion-Israel Institute of Technology, 32000 Haifa, Israel}
\affiliation{Department of Physics, University of Wisconsin, Madison, WI 53706, USA}
\author{Mariana Frank}
\email[]{mfrank@vax2.concordia.ca}
\affiliation{Department of Physics, Concordia University, 7141 Sherbrooke St. 
West, Montreal, Quebec, CANADA H4B 1R6}
\author{Ismail Turan}
\email[]{ituran@physics.concordia.ca}
\affiliation{Department of Physics, Concordia University, 7141 Sherbrooke St.
West, Montreal, Quebec, CANADA H4B 1R6}
\affiliation{Particle Theory Group, Department of Physics,
College of William and Mary, Williamsburg, VA 23187-8795, USA}

\date{\today}

\begin{abstract}
The nonzero widths of heavy particles become significant when they 
appear in the final state of any 
decay occurring just around its kinematical threshold. To take into 
account such effects, a procedure, called the {\it convolution method}, 
was proposed by 
Altarelli, Conti and Lubicz. We expand their study which included only 
threshold effects for $t\to b W Z $ in the standard model.
We discuss finite width effects in the three 
body decays $t\to cWW,cZZ$ and 
$ A^0(h^0)\to t b W$ in  the type III version of a two 
Higgs doublet model. In particular, we find a substantial enhancement in the decay $t\to cZZ$, 
which brings its branching ratio to ${\rm BR}(t\to cZZ)\sim 10^{-3}$, and in the decay 
$A^0\to tbW$, which, unlike the $h^0$ case, becomes competitive with the $A^0$ 
two-body decay modes.
\end{abstract}
\pacs{12.60.-i,12.60.Fr,14.65.Ha,14.80.Bn,14.80.Cp}
\keywords{Finite Width Effects, Rare Top Decays, 
Rare Higgs Decays, Two-Higgs-Doublet-Model}
\vskip -2.5cm
\maketitle
\section{Introduction}\label{}

The finite width of a particle is directly related to its instability. 
When its width is small with respect to its physical mass, 
finite width effects (FWE) are usually neglected except for decays
in which a resonance can emerge when the particle appears as an 
intermediate state, or in decays that are kinematically
allowed only very close to threshold and the particle is involved 
in either the initial or the final state. 
The former case is usually handled with the Breit-Wigner prescription, 
while the latter case, i.e., taking into account the FWE in processes 
occurring just around their kinematical threshold, needs special attention.  

In this respect, there are two different methods proposed in the 
literature \cite{Altarelli:2000nt,Mahlon:1994us,Muta:1986is,Calderon:2001qq}. 
They were referred to by Altarelli, Conti and Lubicz
\cite{Altarelli:2000nt} as 
the {\it decay-chain method} (DCM) and the {\it convolution method} (CM).\footnote{An 
alternative approach has been recently discussed by 
Kuksa \cite{Kuksa:2004cm}, based on the uncertainty relation for the mass of 
the unstable particle. This method has a close analogy to the 
{\it convolution method}.} 
In the first approach (i.e., the DCM), 
the dominant decay modes of the unstable final state  particles are taken 
into account as subsequent decays to obtain the ``total'' decay rate and 
then the branching ratio for the ``signal'' (i.e., with the unstable 
particle in the finite state) is calculated by taking 
the ratio of the ``total'' decay rate to the multiplication of rates 
of the subsequent decay modes \cite{Altarelli:2000nt,Mahlon:1994us}. 
This method requires kinematical cuts 
in order to maintain the direct connection between the ``signal'' and the 
total number of events. That is, since the observed final state 
(with its subsequent decay products) could be 
produced through other channels, kinematical cuts are 
required to minimize this undesired background. Therefore, 
this method leads to physical quantities which depend on kinematical cuts 
and so it inherits some degree of experimental difficulties. 
  
Alternatively, in the 
CM the instability of a final 
state particle is described instead by a Breit-Wigner-like density 
function whose central value and half-width are governed by the width and 
the physical invariant mass of the particle. 
In this way, the unstable particle produced can be seen 
effectively as a real physical particle, 
having an invariant mass which is controlled by its density function. 
Although this method does not require any kinematical cut, 
it doubles the number of phase space integrals, making it 
computationally more challenging. 

In this paper we employ the CM to  
study FWE 
in the three-body flavor changing rare top decays 
$t\to cWW$ and $t\to cZZ$, by including the widths of $W$ and $Z$ bosons. 
These decay modes and other two and three-body rare flavor changing 
top decays \cite{mele}, can provide a unique testing ground for 
the standard model (SM) Glashow-Iliopoulos-Maiani (GIM) mechanism and 
may give hints about - beyond the SM - flavor changing physics 
such as may occur in some variations of Two-Higgs Doublet Models (2HDM's). 
FWE in these decay modes will be studied 
within the SM (in the case of $t\to cWW$) 
and in the context of the type III Two Higgs Doublet Model 
(in both $t\to cWW$ and $t\to cZZ$), 
which admits flavor changing neutral currents (FCNC) at the tree-level. 
The three-body top decays $t\to cWW,bWZ,cZZ$ have been considered before, 
without including FWE, in the SM 
\cite{Jenkins:1996zd,Decker:1992wz,Diaz-Cruz:1999ab},  
in 2HDM's 
\cite{Diaz-Cruz:1999ab,Bar-Shalom:1997tm,Bar-Shalom:1997sj,Li:zv,DiazCruz:1999mq}, in a generic formalism including scalar, vector 
or fermion exchanges \cite{9707229} and in topcolor-assisted Technicolor
model \cite{0103081}. In addition, the top decays 
$t\to bWh^0$ and $t\to bWA^0$ have been analyzed in the context of a
general 2HDM \cite{Iltan:2002am}. 
Among the above decay modes,  
a simple threshold analysis shows that $t\to cZZ$ and $t\to bWZ$ are 
potentially the most sensitive to
FWE. In particular, according to the recent CDF analysis based on the Tevatron 
RUN II data, the top mass is ($1\sigma$) \cite{recenttop1}: 
$m_t=
173.5^{+2.7}_{-2.6}~(stat)\pm 4.0~(syst)$\footnote{Note that
later D0 results from 
Tevatron RUN II, $m_t=170.6\pm 4.2~(stat)\pm 6.0~(syst)$ 
(see \cite{recenttop1}), are based on less accumulated data and has
larger statistical and systematic uncertainties.}
In fact, these later top mass measurements imply that for the stable 
Z-bosons case (i.e., without including FWE) the decay 
$t \to cZZ$ {\it cannot} occur if the top mass lies within its 
recent CDF and D0 $1\sigma$ limits. We, therefore, expect FWE to be 
substantial in this decay. 
Indeed, we find that FWE (due to the rather large ${\mathcal O} (GeV)$ Z-width) 
can give ${\rm BR}(t\to cZZ) \sim 10^{-5} - 10^{-3}$ (as opposed to null 
in the stable case), 
within some range of the allowed parameter space
of the type III 2HDM. Moreover, even for the 
decay $t\to cWW$, for which the central value of the 
top-quark mass (i.e., $m_t=173.5$) is about 10 GeV away from the 
kinematical threshold, we find that FWE from the unstable W-bosons 
can cause a several orders of
magnitudes enhancement in the type III 2HDM with a light 
neutral Higgs of mass $m_{h^0} \lsim 2m_W$, thus elevating the branching ratio 
from ${\rm BR}(t\to cWW) \sim 10^{-9} - 10^{-8}$ 
to ${\rm BR}(t\to cWW) \sim 10^{-4} - 10^{-3}$ in this case.
Clearly, such large branching ratios would be 
accessible to the LHC and may even be detected at the Tevatron.      
A similar large enhancement due to FWE was found 
for the decay mode $t\to bWZ$ in both the CM \cite{Altarelli:2000nt}  
and the DCM \cite{Altarelli:2000nt,Mahlon:1994us}. 
In particular, \cite{Altarelli:2000nt,Mahlon:1994us} have found 
that, in the SM, the 
FWE increase this decay width by orders 
of magnitude (with respect to the stable final state gauge bosons), giving
${\rm BR}(t\to bWZ)\simeq 2\times 10^{-6}$ 
for $m_{t}\sim 176$ GeV.

To demonstrate the potential importance of FWE in neutral Higgs decays, 
we also examine
the three-body neutral Higgs decays $h^0 \to tbW$ and $A^0\to tbW$, 
within the type III 2HDM, assuming that either
$h^0$ (the lighter CP-even neutral Higgs) or $A^0$ 
(the CP-odd neutral Higgs) have masses around $m_t+m_b+m_W$ (i.e., close to 
the threshold).
It is well known that, for a SM-like Higgs, the two-body decay modes  
to the heaviest fermions and to the gauge bosons are dominant, since 
its couplings to these particles are proportional to their masses. 
Three-body sub-threshold decays 
(e.g., to $W^{*}W$ or $Z^{*}Z$ pairs) can also have sizable BR's 
despite the suppression factors involved \cite{Rizzo:1980gz}.
In the context of the minimal supersymmetric extension of the SM (MSSM) 
sub-threshold three-body decays of especially heavy Higgs bosons 
might also have a large branching ratio \cite{Djouadi:1995gv}.      
In this paper we show that, 
including the top quark and the W boson width in the framework 
of the CM, the
three-body Higgs decays $h^0 \to tbW$ and $A^0\to tbW$
can be enhanced by about 3 orders of magnitudes in the type III 
2HDM if they occur just around their kinematical threshold.
For the case of $A^0\to tbW$, such an enhancement 
can push its BR to the level of tens of percents and
may, therefore, become critical for experimental searches of $A^0$.

The paper is organized as follows: In Section II we describe
the {\it convolution method}.  
In Section III we give a brief overview of the
type III 2HDM.
In section IV we examine the FWE in
the top decays $t\to cWW,cZZ$ and in section V
we study the FWE in the three-body Higgs decays 
$h^{0}\to tbW$ and $A^{0}\to tbW$. In Section VI we summarize
our results.

\section{The Convolution Method}\label{}

Particles with large width imply a
large uncertainty in its mass from the mass uncertainty 
relation \cite{Matthews:1958sc}. The CM can be used to include such large width 
effects in decays involving unstable particles in the final state. 
Consider for example the main top decay $t \to bW$. Since the $W$ is unstable, 
we can define: 
$\displaystyle \Gamma(t \to bW) \equiv 
\Gamma=\sum_{i,j}\Gamma^0\left(t\to bf_i\bar{f}_j\right)$, 
where the sum runs over all the $W$ decay modes.  
Furthermore, $\Gamma$ can be decomposed into two parts corresponding to the 
transverse ($\Gamma_T$) and longitudinal ($\Gamma_L$) components of the 
intermediate $W$-boson (see e.g., \cite{Calderon:2001qq}):  
\begin{eqnarray}
\label{transverse}
\Gamma&=&\Gamma_T +\Gamma_L ~,
\end{eqnarray}
\noindent where 
\begin{eqnarray}
\displaystyle
\Gamma_T&=&\frac{1}{\pi}\sum_{ij}\int_{(m_i+m_j)^2}^{(m_t-m_b)^2}dp^2 \frac{\sqrt{p^2}\,\,\,\Gamma^0\Bigl(t\to bW(p^2)\Bigr)\Gamma^0\Bigl(W(p^2)\to f_i\bar{f_j}\Bigr)}{\Bigl(p^2-m_W^2\Bigr)^2+\Bigl(\im \Pi_T(p^2)\Bigr)^2} ~,
\end{eqnarray}
\noindent and $\Gamma_L \propto f(m_i,m_j)$, with $f\to 0$ as $m_i,m_j\to 0$. 
Also, $m_W$ is the mass of the $W$ boson and 
$\im \Pi_T(p^2)$ and $\im \Pi_L(p^2)$ (appearing in $\Gamma_L$) 
are the absorptive parts of the 
transverse and longitudinal vacuum polarization tensor (see e.g., 
\cite{Calderon:2001qq,Atwood:nk}).

Using the Cutkotsky rule 
in the limit of massless fermion $m_i,m_j \to 0$ ($f_i,~f_j$, are
the fermions exchanged in the W self energy diagram), one obtains   
\\$\im\Pi_L(p^2)\to 0$ and: 
\begin{eqnarray}
Im\Pi_T(p^2)=\sqrt{p^2}\sum_{i,j}\Gamma^0\left(W(p^2)\to f_i\bar{f}_j\right) =
\frac{p^2}{m_W}\Gamma_W^0 ~,
\end{eqnarray}
\noindent where $\Gamma_W^0$ is the usual on-shell decay width of $W$ 
and $\sqrt{p^2}\ge m_i+m_j$.
Thus, in this limit $\Gamma$ reduces to:
\begin{eqnarray}
\displaystyle
\label{invariant}
\Gamma=\Gamma_T&=&\int_{0}^{\bigl(m_t-m_b\bigr)^2}dp^2\, \rho\left(p^2,m_W,\Gamma_W^0\right)\Gamma^0\left(t\to bW(p^2)\right) ~,
\end{eqnarray}
\noindent where $\rho\left(p^2,m_W,\Gamma_W^0\right)$ is the 
``invariant mass distribution function'', given by: 
\begin{eqnarray}
\rho\left(p^2,m_W,\Gamma_W^0\right)&=&\frac{1}{\pi}\frac{\frac{p^2}{m_W}\Gamma_W^0}{\Bigl(p^2-m_W^2\Bigr)^2+ 
\left(\frac{p^2}{m_W}\Gamma_W^0\right)^2} \label{rho}~.
\end{eqnarray}
\noindent Eqs.~(\ref{invariant}) and (\ref{rho}) describe the 
factorization of the production and the 
decay modes of the $W$ boson (in the limit of massless fermions). 
The case of a stable $W$ boson (i.e., $\Gamma_W^0\to 0$) makes 
$\rho\to\delta(p^2-m_W^2)$ which sets 
$\Gamma=\Gamma^0(t\to bW)$, where $\Gamma^0$ is the width 
for an on-shell $W$ without FWE.

The above prescription can be generalized to the case of a generic  
three-body decay of the form $a\to bV_1V_2$, where 
$V_1$ and $V_2$ are vector bosons:
\begin{eqnarray}
\label{twoinvariant}
\!\!\!\!\!\Gamma(a\to bV_1V_2)=\int_{0}^{\bigl(m_a-m_b\bigr)^2}\!\!\!\!\!\!dp_1^2\int_{0}^{\bigl(m_a-m_b-\sqrt{p_1^2}\bigr)^2}
\!\!dp_2^2&&\rho_1\left(p_1^2,m_{V_1},\Gamma_{V_1}^0\right)
\rho_2\left(p_2^2,m_{V_2},\Gamma_{V_2}^0\right)\nonumber\\
&&\times\Gamma^0\left(a\to bV_1(p_1^2)V_2(p_2^2)\right) \label{v1v2}.
\end{eqnarray}
\noindent Furthermore, for consistency of the CM one needs 
also the following modifications:
\begin{enumerate} 
\item The sum over polarization vectors of a gauge-boson with 
an invariant mass $p^2$ should be taken as:
\begin{eqnarray}
\sum_{\lambda}\epsilon_{\lambda}^{\mu}(p)\epsilon_{\lambda}^{\nu *}(p)=-g^{\mu\nu}+\frac{p^{\mu}p^{\nu}}{p^2}~.
\end{eqnarray}
\item In calculating the ``zeroth'' width of the top-quark(Higgs) into the off-shell 
vector boson(s) [i.e., $\Gamma^0\left(a\to bV_1(p_1^2)V_2(p_2^2)\right)$ in 
Eq.~(\ref{v1v2}) or $\Gamma^0\left({\cal H}\to a\,b(p_1)V(p_2^2)\right)$ for ${\cal H}=h^0$ in Eq.~(\ref{tV})], the 
tree-level propagator of the massive vector bosons should be modified as 
(in the unitary gauge):
\begin{eqnarray}
\frac{-i}{p^2-m_{V}^2+im_{V}\Gamma_{V}^0}\left[g^{\mu\nu}-\frac{p^{\mu}p^{\nu}}{m_{V}^2-im_{V}\Gamma_{V}^0}\right]\label{Vectorprop} ~.
\end{eqnarray}
\noindent This substitution is required since in the CM
the invariant mass $p^2$ is allowed to vanish, as can be seen from the 
integration limits in 
Eqs.~(\ref{invariant}) and (\ref{v1v2}).

\item In order to restore gauge invariance in the $R_{\xi}$-gauge, 
the Feynman rules in which masses of such resonant intermediate particles 
appear should be modified to be functions of the corresponding 
invariant masses. Such a 
modification is, however, not necessary in the unitary gauge that we 
have used in all our calculations \cite{LopezCastro:1991nt}. 

\end{enumerate}

\section{The two Higgs doublet model of type III}\label{}

One of the simplest extensions of the SM is obtained by enlarging the scalar 
sector with an additional $SU(2)_L$ doublet. 
In the most general case such a 2HDM gives rise to tree-level 
FCNC which are mediated by the physical Higgs bosons 
\cite{luke}.
To avoid such potentially dangerous FCNC, one usually imposes 
an ad-hoc discrete symmetry \cite{Glashow:1976nt} that leads to 
the type I or type II 2HDM (see for example \cite{HHG} and \cite{ourreview}). 
An alternative way for suppressing FCNC
in a general 2HDM (i.e., without imposing discrete symmetries) 
was suggested by Cheng and Sher in \cite{Cheng:1987rs}. 
In the Cheng and Sher Ansatz the arbitrary 
flavor changing couplings of the scalars to fermions are assumed 
to be 
proportional to the square root of masses of the fermions 
participating in the Higgs Yukawa vertex (see below).\footnote{The Cheng and 
Sher Ansatz ensures the suppression of
FCNC within the first two generations of quarks, as required by 
the experimental constraints on FCNC in meson transitions, 
see \cite{Atwood:1996vj}.} 

Within the most general 2HDM one can always choose a basis
where only one of the doublets acquires a vacuum expectation value (VEV):
$\langle \Phi_1\rangle=\left(0\,\,\,\,v/\sqrt{2}\right)^T\,\, \rm{and}\,\,
\langle\Phi_2\rangle=0$. A general 2HDM in this basis is often referred to as 
the type III 2HDM (or Model III) 
\cite{Atwood:1996vj,Sher:1991km,Atwood:1995ej}. 
With this choice of basis, $\Phi_1$ corresponds to the usual SM doublet 
and all the new flavor changing couplings are attributed to $\Phi_2$. 
Note also that in this basis $\tan\beta=v_1/v_2$ has no physical 
meaning.\footnote{``Switching on'' $\tan\beta$ by allowing 
$\langle\Phi_2\rangle \neq 0$ will not change 
any physical result.}    

As in any 2HDM, the physical Higgs sector of Model III consists of 3 
neutral Higgs bosons (2 CP-even ones, $h^0$ and $H^0$, and one CP-odd state
$A^0$) and a charged scalar with its conjugate $H^{\pm}$. The neutral bosons are 
given, in terms of the original SU(2) doublets, as:   
\begin{eqnarray}
h^0&=&\sqrt{2}\Bigl[-\Bigl(\re \phi_1^0-v\Bigr)\,\sin\alpha+\re \phi_2^0\, \cos\alpha\Bigr]\,,\nonumber\\
H^0&=&\sqrt{2}\Bigl[\Bigl(\re \phi_1^0-v\Bigr)\,\cos\alpha+\re \phi_2^0\, \sin\alpha\Bigr]\,,\nonumber\\
A^0&=&-\sqrt{2}\im \phi_2^0 ~.
\end{eqnarray}
\noindent The flavor changing part of the Yukawa Lagrangian in Model III 
is given by \cite{luke,Atwood:1996vj}:
\begin{eqnarray}
{\mathcal L}_{Y,FC}=\xi_{ij}^U \bar{Q}_{iL}\tilde{\phi}_2U_{jR}+\xi_{ij}^D \bar{Q}_{iL}\phi_2D_{jR}+{\rm H.c.} ~,
\end{eqnarray}
\noindent where $\tilde{\phi}_2=i\tau_2\phi_2$, 
$Q$ stands for the quark $SU(2)_L$ doublets, 
$U(D)$ for up-type (down-type) quark $SU(2)_L$ singlets 
and $\xi^U,~\xi^D$ are $3\times3$ non-diagonal matrices (in family space) 
that parametrize the strength of the FCNC vertices in the neutral Higgs 
sector. Adopting the Cheng and Sher Ansatz we set:\footnote{Note that 
there is a factor of 1/2 difference between our definition 
for $\xi_{ij}^{U,D}$ in (Eq.~\ref{xiud}) and the one used 
in \cite{Bar-Shalom:1997sj}. This difference may be absorbed by 
redefining the arbitrary parameters $\lambda_{ij}$.}
\begin{eqnarray}
\xi_{ij}^{U,D}=\lambda_{ij}\frac{\sqrt{m_i m_j}}{v}\,,\,\,\,v=\left(\sqrt{2}G_F\right)^{-1/2} \label{xiud} ~,
\end{eqnarray}
\noindent where for simplicity we assume the 
$\lambda_{ij}$'s to be real\footnote{In this work we are not 
interested in CP-violating effects that may be driven by a possible phase 
contained in the $\lambda_{ij}$'s.} 
and symmetric (i.e., $\lambda_{ij}^*=\lambda_{ji}$) constants. 
For the Higgs-top-charm coupling we will take that 
$\lambda_{tc}=\lambda_{ct} \equiv \lambda \sim {\mathcal O}(1)$, which 
is compatible with all existing data, see \cite{Bar-Shalom:1997sj,Atwood:1996vj}
for details.

Thus, for the top decays of our interest in this paper, 
the relevant terms in the Yukawa Lagrangian are \cite{Bar-Shalom:1997sj}:
\begin{eqnarray}
{\mathcal L}_{{\mathcal H}tc}&=&-\lambda \frac{\sqrt{m_c m_t}}{\sqrt{2}v}f_{{\mathcal H}} {\mathcal H} \bar{c}t\,,\nonumber\\
{\mathcal L}_{{\mathcal H}VV}&=&- g m_W G_V S_{{\mathcal H}}g_{\mu\nu}V^{\mu}V^{\nu}\,,
\end{eqnarray}
\noindent where ${\mathcal H}=h^0$ or $H^0$,  $V=W$ or $Z$ and 
\begin{eqnarray}
f_{h^0;H^0}&=&\cos\alpha;\,\,\sin\alpha\,,\nonumber\\
S_{h^0;H^0}&=&\sin\alpha;\,\,-\cos\alpha\,,\nonumber\\
G_{W;Z}&=&1;\,\,\frac{m_Z^2}{m_W^2} ~.
\end{eqnarray}
\noindent We will further need the ${\mathcal H}q_iq_i$ (with ${\mathcal H}=h^0,A^0$)
and $H^{\pm}tb$ couplings \cite{Bar-Shalom:1997sj,Atwood:1996vj}:
\begin{eqnarray}
{\mathcal L}_{{\mathcal H}q_iq_i}&=&-\frac{m_{q_i}}{v}\bar{q}_i\left[h^0\left(-\sin\alpha +\frac{\lambda_{ii}}{\sqrt{2}}\right)+iA^0\frac{\lambda_{ii}}{\sqrt{2}}\gamma_5\right]q_i\,,\nonumber\\
{\mathcal L}_{H^-t\bar{b}}&=&-\frac{1}{2v}V_{tb}^*H^-\bar{b}\Biggl[\biggl(\lambda_{bb}m_b-\lambda_{tt}m_t\biggr)-\biggr(\lambda_{bb}m_b+\lambda_{tt}m_t\biggr)\gamma_5\Biggr]t 
\label{yukawa}~.
\end{eqnarray}

\section{Finite width effects in the $t\to cWW$ and $t\to cZZ$ decays}\label{}

In this section we will use the CM to evaluate the FWE in the top
decays $t\to cWW$ and $t\to cZZ$.
Kinematically, 
the naive threshold (i.e., not including FWE) 
for the decay $t\to cZZ$ is about 4 GeV away (i.e., larger) from
the recent CDF $1\sigma$ limit (from Tevatron RUN II) on the
top mass, $m_t(1\sigma) \leq  180.2$ GeV \cite{recenttop1}.
Also, as will be shown below,
even for $t\to cWW$ the available phase space can be 
(depending on the top mass) small enough for the FWE to become significant.

We will consider the decay $t\to cWW$ at the tree-level in both the SM and 
Model III, while $t\to cZZ$ will be analysed only within Model III, since
in the SM this decay is doubly suppressed by both
one-loop factors and non-diagonal Cabibbo-Kobayashi-Maskawa (CKM) elements
and is, therefore, unobservably small. 

In the SM, the tree-level decay $t\to cWW$ proceeds
via $t\to d^* W^+ \to c W^-W^+$ ($d=d,s$ or $b$ quarks),
with a BR of the order of ${\mathcal O}(10^{-14}-10^{-13})$ (depending 
on the top-quark mass) if FWE 
are not taken into account \cite{Jenkins:1996zd}. 
The dominant SM diagram is $t\to b^*W^+ \to c W^-W^+$, since
$V_{tb} \times V_{cb}$ is the largest out of the three possible
products of CKM elements that enter 
this decay. 
In Model III there are two additional tree-level diagrams: 
$t\to ch^{0*}\to cW^+W^-$ and $t\to cH^{0*}\to cW^+W^-$ 
\cite{Bar-Shalom:1997tm,Bar-Shalom:1997sj}. 
In this case, we will use 
the Breit-Wigner prescription for the propagators of 
${\mathcal H}=h^0$~or~$H^0$, i.e.,
$(q^2-m_{\mathcal H}^2+im_{\mathcal H}\Gamma_{\mathcal H})^{-1}$, 
where $\Gamma_{\mathcal H}$ is the total ${\mathcal H}$ width calculated 
from the dominant ${\mathcal H}$ decay modes: 
${\mathcal H}\to b\bar{b},t\bar{t},t\bar{c},ZZ,WW,WW^*,ZZ^*$.\footnote{Note that 
in Model III the decay ${\mathcal H}\to t\bar{c}$ becomes important for
$\lambda_{tc} \sim {\mathcal O}(1)$.}$^,$\footnote{
Depending on the ${\mathcal H}$ mass, 
only the kinematically allowed decays will be included 
in $\Gamma_{\mathcal H}$.} 

Using the CM, the partial decay 
width for $t\to cWW$ in any given model $M$ can be written as
[see Eq.~(\ref{twoinvariant})]:
\begin{eqnarray}
\displaystyle
\!\!\!\!\!\!\!\!\!\!\!\!\!\!\!\!\Gamma^M_{\rm conv}(t\rightarrow cWW)=&&\!\!
\frac{1}{512 \pi^3 m_{t}^3}\!\!\int_{0}^{\bigl(m_{t}-m_{c}\bigr)^2}\!\!dp_{W^+}^2 
\left[\frac{p_{W^{+}}^2 \Gamma_{W}^0}{m_{W}\pi \left( \Bigl(p_{W^{+}}^2-m_{W}^2\Bigr)^2+
\Bigl(\frac{p_{W^{+}}^2 \Gamma_{W}^0}{m_{W}}\Bigr)^2\right)}\right]\nonumber\\
&&\!\!\!\times\int_{0}^{\bigl(m_{t}-m_{c}-\sqrt{p_{W^+}^2}\bigr)^2}\!\!dp_{W^-}^2
\left[\frac{p_{W^{-}}^2 \Gamma_{W}^0}{m_{W}\pi \left( \Bigl(p_{W^{-}}^2-m_{W}^2\Bigr)^2+
\Bigl(\frac{p_{W^{-}}^2 
\Gamma_{W}^0}{m_{W}}\Bigr)^2 \right)}\right]\nonumber\\
&&\!\!\!\times\int_{\bigl(m_{c}+\sqrt{p_{W^-}^2}\bigr)^2}^{\bigl(m_{t}-
\sqrt{p_{W^+}^2}\bigr)^2}\!\!dx_{1}\int_{x_{2,min}}^{x_{2,max}}\!dx_{2}\,\,
\left|{\mathcal M}^M_{\rm conv}(x_{1},x_{2},p_{W^+}^2,p_{W^-}^2)\right|^2\!,
\label{SMgamma} 
\end{eqnarray}
\noindent where the superscript $M$ stands for the model used for the calculation of the convoluted amplitude ${\mathcal M}^M_{\rm conv}$, and 
\begin{eqnarray}
x_{2,min}&=&(E_{2}+E_{3})^{2}-\bigg(\sqrt{E_{2}^{2}-p_{W^{-}}^{2}}+\sqrt{E_{3}^{2}-p_{W^{+}}^{2}}\bigg)^{2} ~,\nonumber\\
x_{2,max}&=&(E_{2}+E_{3})^{2}-\bigg(\sqrt{E_{2}^{2}-p_{W^{-}}^{2}}-\sqrt{E_{3}^{2}-p_{W^{+}}^{2}}\bigg)^{2}~,\nonumber\\
E_{2}&=&\frac{x_{1}-m_{c}^2+p_{W^{-}}^2}{2\sqrt{x_1}};\;\;E_{3}=\frac{-x_{1}-p_{W^{+}}^2+m_{t}^2}{2\sqrt{x_1}}~.
\end{eqnarray}
\noindent For the BR calculation, we approximate the total width 
of the top quark by its dominant decay $t\to bW$ which is computed at 
tree-level with the corresponding value of the top quark mass.

\begin{figure}[htb]
\vspace{-2.8in} 
    \centerline{ \epsfxsize 6.0in {\epsfbox{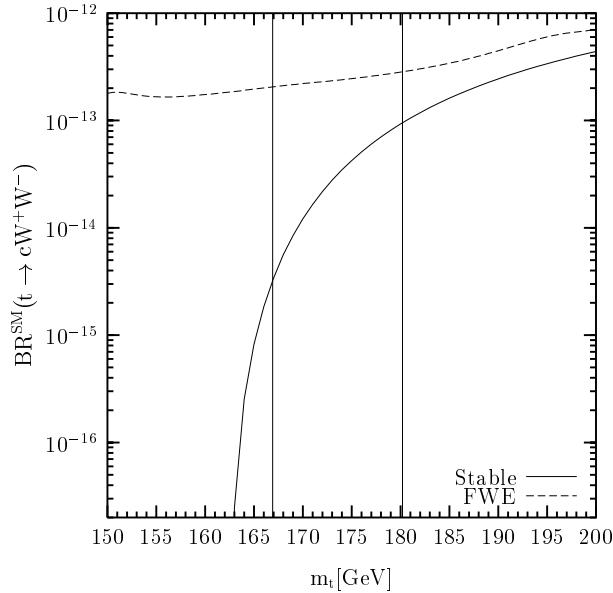}} }
\vskip -2.3in
    \caption{\emph {The branching ratio 
for $t\rightarrow cWW$ in the SM, 
as a function of the top quark mass, without FWE (solid curve) and 
with FWE (dashed curve). The charm quark mass is set to $m_c=1.87$ GeV and 
$V_{cb}=0.046$. The vertical lines denote the recent 
(from Tevatron RUN II) 
lower and upper  
$1\sigma$ CDF limits on the top mass.}}\label{figtcWW1}
\end{figure}

In Fig.~\ref{figtcWW1} we plot the ${\rm BR}(t \to c W^+W^-)$ as a function
of the top quark mass in the SM, with and without FWE. The case of 
stable $W$'s in the final state (i.e. without FWE) is obtained 
by taking the limit 
$\rho(p_W^2,m_W^2,\Gamma_W^0) \to \delta(p_W^2-m_W^2)$ [see Eq.~(\ref{rho})] 
which sets $p_{W^{\pm}}^2=m_W^2$ in the integrand of Eq.~(\ref{SMgamma}). 
The decay $t \to c W^+W^-$ in the SM with stable $W$'s 
was calculated in \cite{Jenkins:1996zd} and our result for this case agrees with 
hers.
From Fig.~\ref{figtcWW1}
we see that for the CDF central value of the top mass, $m_t =173.5$ GeV,  
FWE can enhance the ${\rm BR}(t \to c W^+W^-)$ by about an order of magnitude,
reaching $\sim 2 \cdot 10^{-13}$. For the lower $1\sigma$ CDF limit 
$m_t \sim 167$ GeV, the enhancement due to FWE is of about two 
orders of magnitudes.   
Unfortunately, even with such large FWE in the decay 
$t \to c W^+W^-$, the BR in the SM 
is still too small to be measured - even at the LHC.
%
%
\begin{figure}[htb]
\vspace{-2.8in}  
    \centerline{ \epsfxsize 5.5in {\epsfbox{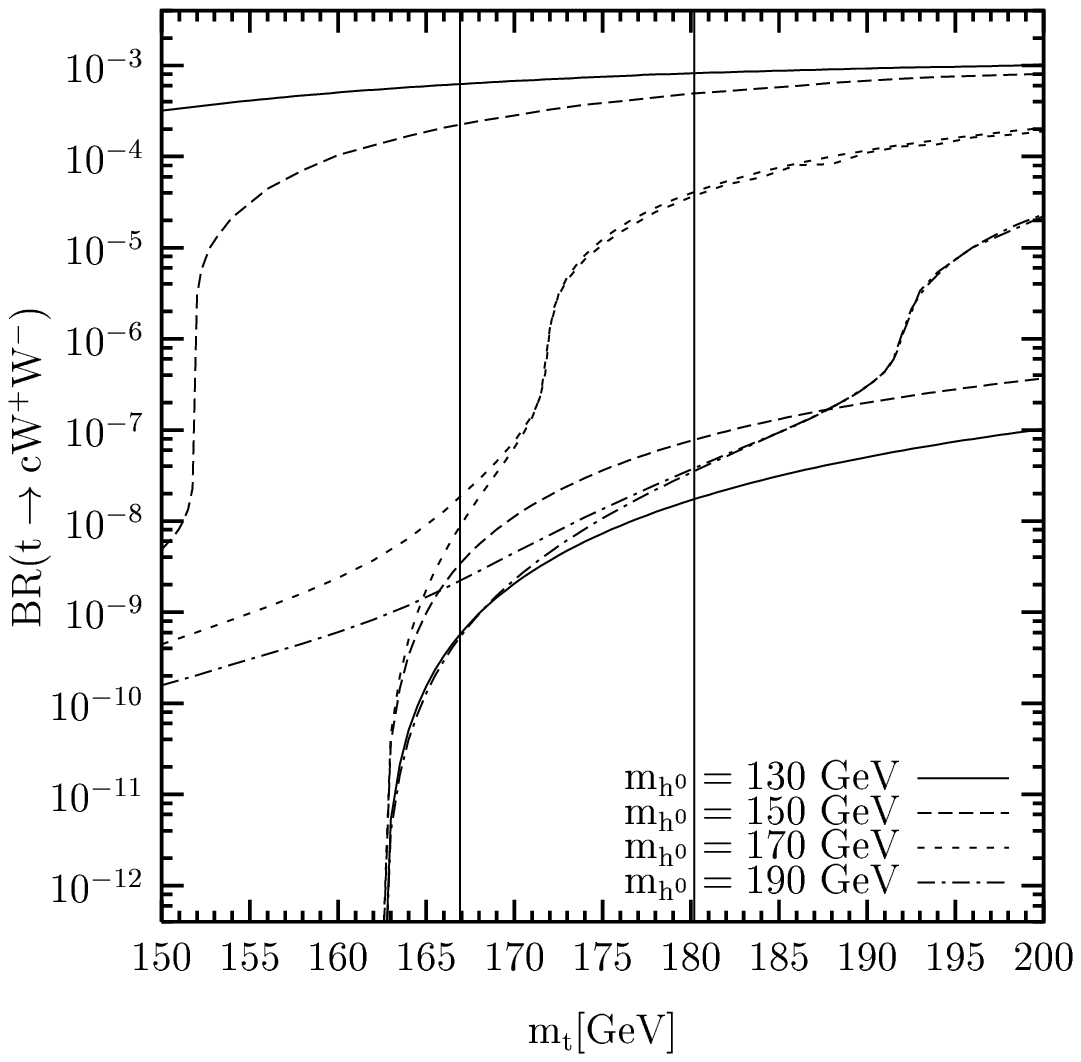}} \hspace{-7cm} \epsfxsize 5.5in {\epsfbox{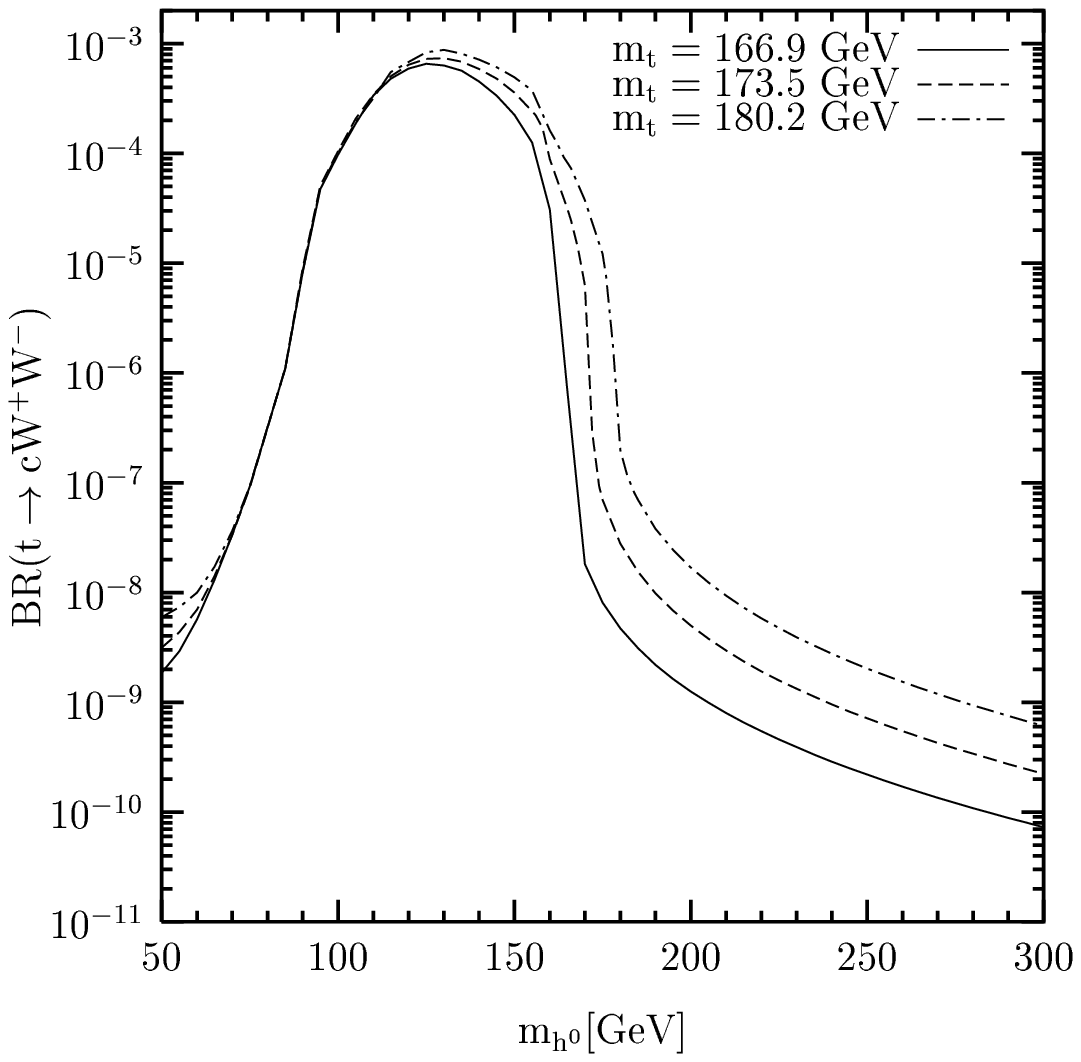}} }
\vskip -2.2in
    \caption{\emph {The branching ratio for $t\rightarrow cWW$ in Model III, 
for $\lambda_{tc}=1$, $m_{H^0}=1$ TeV and $\alpha=\pi/4$. On the left: 
as a function of the top quark mass, without FWE (lower curves) and 
with FWE (upper curves), for 
$m_{h^0}=130,~150,~170,~190$ GeV. On the right: as a function of 
the light Higgs mass $m_{h^0}$, 
for $m_{t}=166.9,~173.5,~180.2$ GeV. See also caption to
Fig.~\ref{figtcWW1}.}}\label{figtcWW2}
\end{figure}

In Fig.~\ref{figtcWW2} we show the  
${\rm BR}(t \to c W^+W^-)$ in Model III with 
$\lambda_{tc}=1$, $m_{H^0}=1$ TeV and $\alpha=\frac{\pi}{4}$\,\footnote{The dependence
of ${\rm BR}(t \to c W^+W^-)$ and ${\rm BR}(t \to c ZZ)$ on the Higgs mixing angle 
$\alpha$ in Model III can 
be found in \cite{Bar-Shalom:1997tm,Bar-Shalom:1997sj}. 
The maximum of these branching ratios with respect to  
$\alpha$ takes place at 
$\alpha=\pi/8$ (due to the dependency of the Higgs width on 
$\alpha$) and not at $\alpha=\pi/4$ which is used through out our analysis.}  
(note that the SM tree-level 
contribution to $t \to c WW$, although included, is negligible in this case), 
as a function
of $m_t$ with and without FWE, for several values of the light Higgs mass 
$m_{h^0}=130,~150,~170,~ {\rm and}~ 190$ GeV, and as a function 
of $m_{h^0}$ with FWE, for the lower, upper and central CDF 
values of the top-quark mass $m_t=166.9~,173.5$, and 180.2 
GeV.
As was found in \cite{Bar-Shalom:1997tm,Bar-Shalom:1997sj}, 
in Model III without FWE, the ${\rm BR}(t \to c W^+W^-)$ 
can at most reach the level of ${\rm few} \times 10^{-5}$ 
if $m_t$ lies within its $1\sigma$ CDF limits and {\it only if} 
$m_{h^0} \sim m_t$. 
On the other hand, 
when FWE are ``turned on'', a huge enhancement to the width
arises within a large range of the Higgs mass.
In particular, for $ 100 ~{\rm GeV} \lsim m_{h^0} \lsim 165$ GeV, we find 
${\rm BR}(t \to c W^+W^-) \gsim 10^{-4}$, if   
$ 167 ~{\rm GeV} \lsim m_t \lsim 180$ GeV, 
in Model III when FWE are included.
Note that, for the lower $1\sigma$ limit $m_t \sim 167$ GeV, i.e.,
close to the threshold for producing $cWW$, the FWE causes an up to 
six orders of magnitudes enhancement to the ${\rm BR}(t \to c W^+W^-)$ if, 
e.g., $m_h \sim 130$ GeV.

\begin{figure}[htb] 
\vspace{-2.6in}
    \centerline{ \epsfxsize 5.5in {\epsfbox{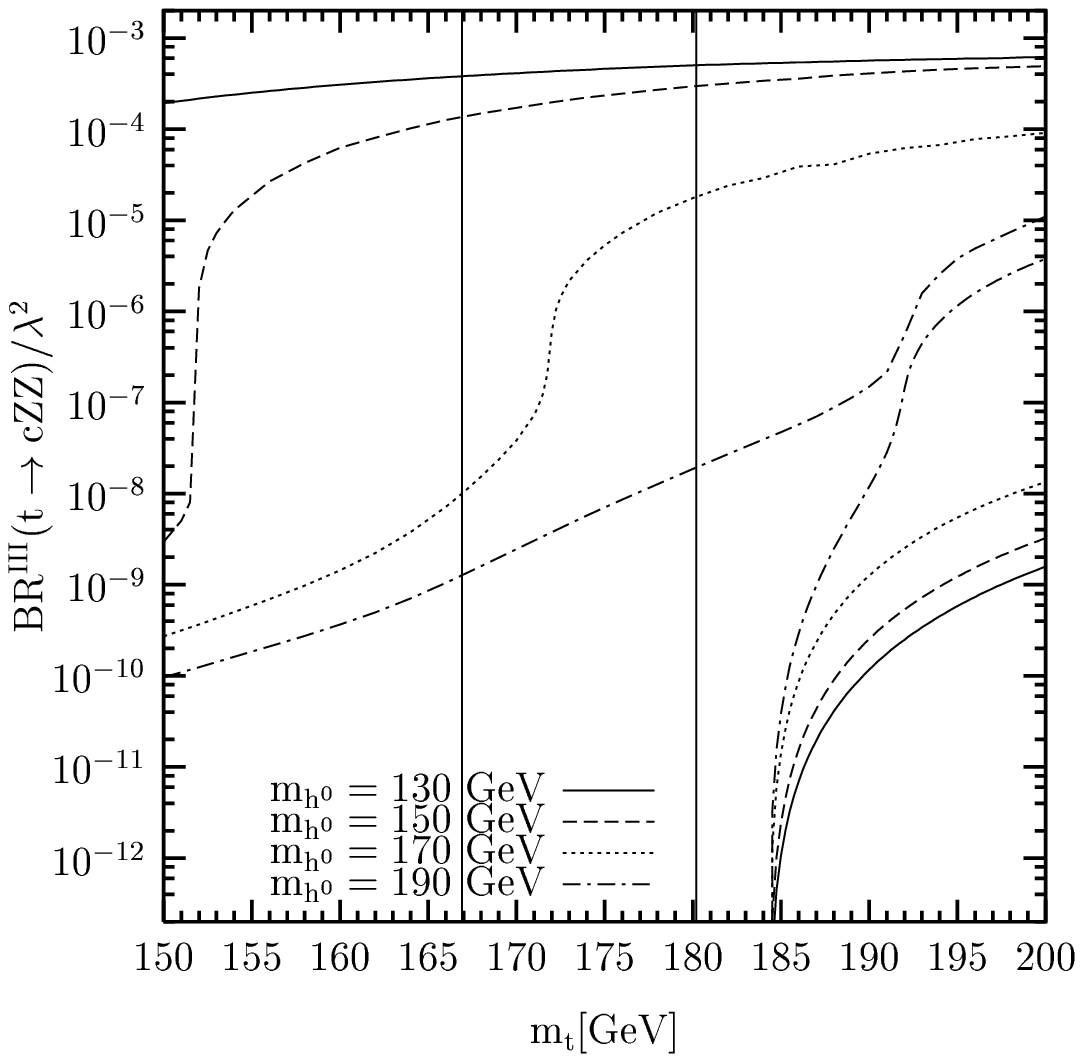}} \hspace{-7cm} \epsfxsize 5.5in {\epsfbox{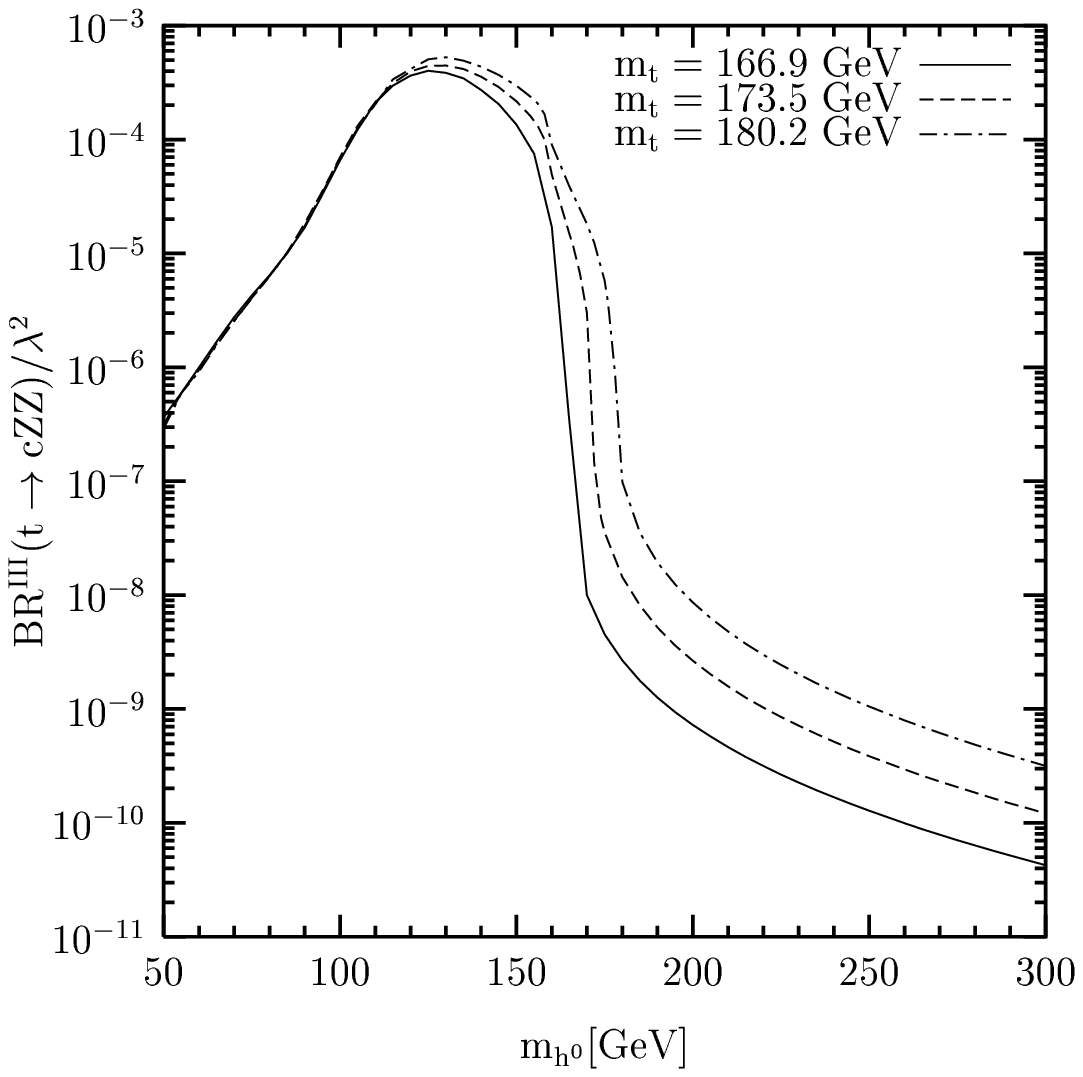}} }
\vskip -2.2in
    \caption{\emph {The 
scaled branching ratio ${\rm BR}(t\to cZZ)/\lambda^2$ in Model III, 
for $m_{H^0}=1$ TeV and $\alpha=\pi/4$. On the left: 
as a function of the top quark mass, without FWE (lower curves) and 
with FWE (upper curves), for 
$m_{h^0}=130,~150,~170,~190$ GeV. On the right: as a function of 
the light Higgs mass $m_{h^0}$, 
for $m_{t}=166.9,~173.5,~180.2$ GeV. See also caption to
Fig.~\ref{figtcWW2}.}}\label{figtcZZ1}
\end{figure}

For the decay $t \to c ZZ$ in Model III 
we use 
the analytical results of $t\to cWW$ with the replacements
$m_W\to m_Z/\cos\theta_W$ in the ${\mathcal H} VV$ vertex,
$p_W^-\to p_{Z_1},\,p_W^+\to p_{Z_2}$ in Eq.~(\ref{SMgamma}) and 
with an additional overall factor of 1/2 to take into account 
the symmetry factor for identical particles in the final state 
(i.e., $Z$ bosons). 
Fig.~\ref{figtcZZ1} shows the 
scaled branching ratio ${\rm BR}(t\to cZZ)/\lambda^2$ 
($\lambda \equiv \lambda_{tc}$) in Model III 
with $m_{H^0}=1$ TeV and $\alpha=\pi/4$ 
(see also footnote 9),
as a function
of $m_t$ with and without FWE, for  
$m_{h^0}=130,~150,~170,~{\rm and}~190$ GeV, and as a function 
of $m_{h^0}$ with FWE, for $m_t=166.9~,173.5$, and 180.2 
GeV.
Note that the decay $t\to cZZ$ is fundamentally different from 
$t\to cWW$, since, unlike $t\to cWW$, this decay channel 
{\it cannot} occur for stable Z-bosons 
if $m_t$ lies within its $1\sigma$ limits. Thus, 
the inclusion of FWE in $t\to cZZ$ is crucial in this case. 
In particular,
from Fig.~\ref{figtcZZ1} we see
that a remarkably large ${\rm BR}(t\to cZZ) \sim 10^{-5} - 10^{-3}$ 
is expected in Model III, if $m_{h^0}$ lies within 
$ 90 ~{\rm GeV} \lsim m_{h^0} \lsim 170$ GeV. Such a large BR will 
be accessible to the LHC and may even be detected at the 
Tevatron.   

Finally we note that, following \cite{Altarelli:2000nt} (who took 
$m_b=m_B$ for their calculation of $t\to bWZ$), 
we take $m_c=m_D=1.87$ GeV.
%
\section{Finite width effects in the $A^0\to(\bar{t}b+t\bar{b})W$ and 
$h^0\to(\bar{t}b+t\bar{b})W$ decays}\label{}

In this section we will examine FWE in three-body decays of neutral Higgs 
bosons in Model III. We will focus on the decay channels 
$A^{0} \to \bar{t}bW^{+}$ and $h^0\to \bar{t}bW^{+}$ which can 
have both theoretical and 
experimental advantages for Higgs searches and for investigating 
Higgs properties in the Higgs mass range 
$200 ~{\rm GeV} \lsim m_{h^0},~m_{A^0} \lsim 300$ GeV.  

\begin{figure}[htb]
\begin{center}
\begin{picture}(200,90)(10,-5)
\hspace*{1cm}
\SetWidth{1.0}
\Text(0,50)[c]{$(a)$}
\DashArrowLine(0,0)(50,0){2}
\ArrowLine(50,50)(50,0)
\ArrowLine(50,0)(100,0)
\Photon(100,0)(150,0){3}{6}
\ArrowLine(50,50)(50,0)
\ArrowLine(100,0)(100,50)
\ArrowLine(55,30)(55,50)
\Text(25,10)[c]{${A^{0}(h^{0})}$}
\Text(67,40)[c]{$p_{t}$}
\Text(40,45)[c]{$\bar{t}$}
\Text(75,10)[c]{$t$}
\Text(105,45)[c]{$b$}
\Text(75,10)[c]{$t$}
\Text(145,13)[c]{${W_{\mu}^+}$}
\Vertex(50,0){2}
\Vertex(100,0){2}
\end{picture}
\begin{picture}(200,90)(130,0)
\SetWidth{0.8}
\Text(150,50)[c]{$(b)$}
\DashArrowLine(150,0)(200,0){2}
\DashArrowLine(200,0)(250,0){2}
\ArrowLine(250,0)(300,0)
\ArrowLine(250,50)(250,0)
\Photon(200,50)(200,0){3}{6}
\ArrowLine(245,30)(245,50)
\Text(175,10)[c]{${A^0(h^0)}$}
\Text(228,10)[c]{${H^-}$}
\Text(275,10)[c]{$b$}
\Text(260,45)[r]{$\bar{t}$}
\Text(228,40)[l]{$p_{t}\!\!$}
\Text(175,45)[l]{${W^+}$}
\Vertex(200,0){2}
\Vertex(250,0){2}
\end{picture}
\begin{picture}(200,90)(10,-5)
\hspace*{1cm}
\SetWidth{1.0}
\Text(0,50)[c]{$(c)$}
\DashArrowLine(0,0)(50,0){2}
\ArrowLine(50,0)(50,50)
\ArrowLine(100,0)(50,0)
\Photon(100,0)(150,0){3}{6}
\ArrowLine(100,50)(100,0)
\ArrowLine(95,30)(95,50)
\Text(25,10)[c]{${A^{0}(h^{0})}$}
\Text(44,45)[c]{$b$}
\Text(108,45)[c]{$\bar{t}$}
\Text(78,10)[r]{$b$}
\Text(78,40)[l]{$p_{t}$}
\Text(145,13)[c]{${W_{\mu}^+}$}
\Vertex(50,0){2}
\Vertex(100,0){2}
\end{picture}
\begin{picture}(200,90)(130,-5)
\SetWidth{1.0}
\Text(150,50)[c]{$(d)$}
\DashArrowLine(150,0)(200,0){2}
\Photon(200,0)(250,0){3}{6}
\ArrowLine(250,0)(300,0)
\ArrowLine(250,50)(250,0)
\Photon(200,0)(200,50){3}{6}
\ArrowLine(245,30)(245,50)
\Text(175,10)[c]{${h^0}$}
\Text(228,10)[c]{${W^-}$}
\Text(275,10)[c]{$b$}
\Text(260,45)[r]{$\bar{t}$}
\Text(228,40)[l]{$p_{t}$}
\Text(170,45)[l]{${W^+}$}
\Vertex(200,0){2}
\Vertex(250,0){2}
\end{picture}
\end{center}
\vspace*{-0.5cm}
\caption{\emph{Tree-level diagrams contributing to the decay 
$A^{0}(h^0)\to \bar{t}bW^{+}$ in Model III.}}\label{Feynm}
\caption{\emph{Tree-level diagrams contributing to the decay 
$A^{0}(h^0)\to \bar{t}bW^{+}$ in Model III.}}\label{Feynm}
\end{figure}

The tree level diagrams contributing to these two decays in Model III 
are given in Fig. \ref{Feynm} (note that, for the $A^0$ decay, the diagram 
with an intermediate $W$-boson is missing, i.e., diagram (d), due to 
the absence of a tree-level $A^0WW$ coupling).
A fomula analogous to Eq.~(\ref{v1v2}) can be given for Higgs decays as
\begin{eqnarray}
\!\!\!\!\!\Gamma({\cal H}\to b\, \bar{a}\, V)=\int_{0}^{\bigl(m_{\cal H}-m_b\bigr)^2}\!\!\!\!\!\!dp_1^2\int_{0}^{\bigl(m_{\cal H}-m_b-\sqrt{p_1^2}\bigr)^2}
\!\!dp_2^2&&\rho_1\left(p_1^2,m_t,\Gamma_{a}^0\right)
\rho_2\left(p_2^2,m_V,\Gamma_V^0\right)\nonumber\\
&&\times\Gamma^0\left({\cal H}\to b\, \bar{a}(p_1^2)\, V(p_2^2))\right) \label{tV},
\end{eqnarray}
where ${\cal H}=h^0$ or $A^0$ and $a(b)$ is the top(bottom) quark.
Using the interaction terms in Section 3, 
we calculate the matrix element for 
each decay, where: 
\begin{figure}[htb]
\vspace{-2.7in} 
    \centerline{ \epsfxsize 5.8in {\epsfbox{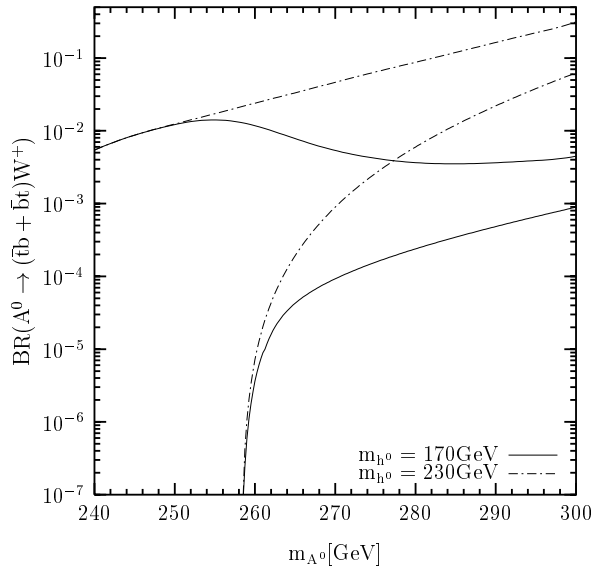 }} }
\vskip -2.3in
    \caption{\emph {The branching ratio
${\rm BR}(A^0 \to t \bar b W^- + \bar t b W^+)$ in model III,
as a function of the $A^0$ mass, 
with FWE (upper curves) and without FWE (lower curves), for 
$m_{h^0}=170$ GeV (solid curves) and
$m_{h^0}=230$ GeV (dashed-dotted curves). Also,  
$m_t=173.5$ GeV, $m_{H^+}=350$ GeV 
and $\alpha=\pi/4$.}}\label{BRA0}
\end{figure}

\begin{figure}[htb]
\vspace{-2.7in} 
    \centerline{ \epsfxsize 5.8in {\epsfbox{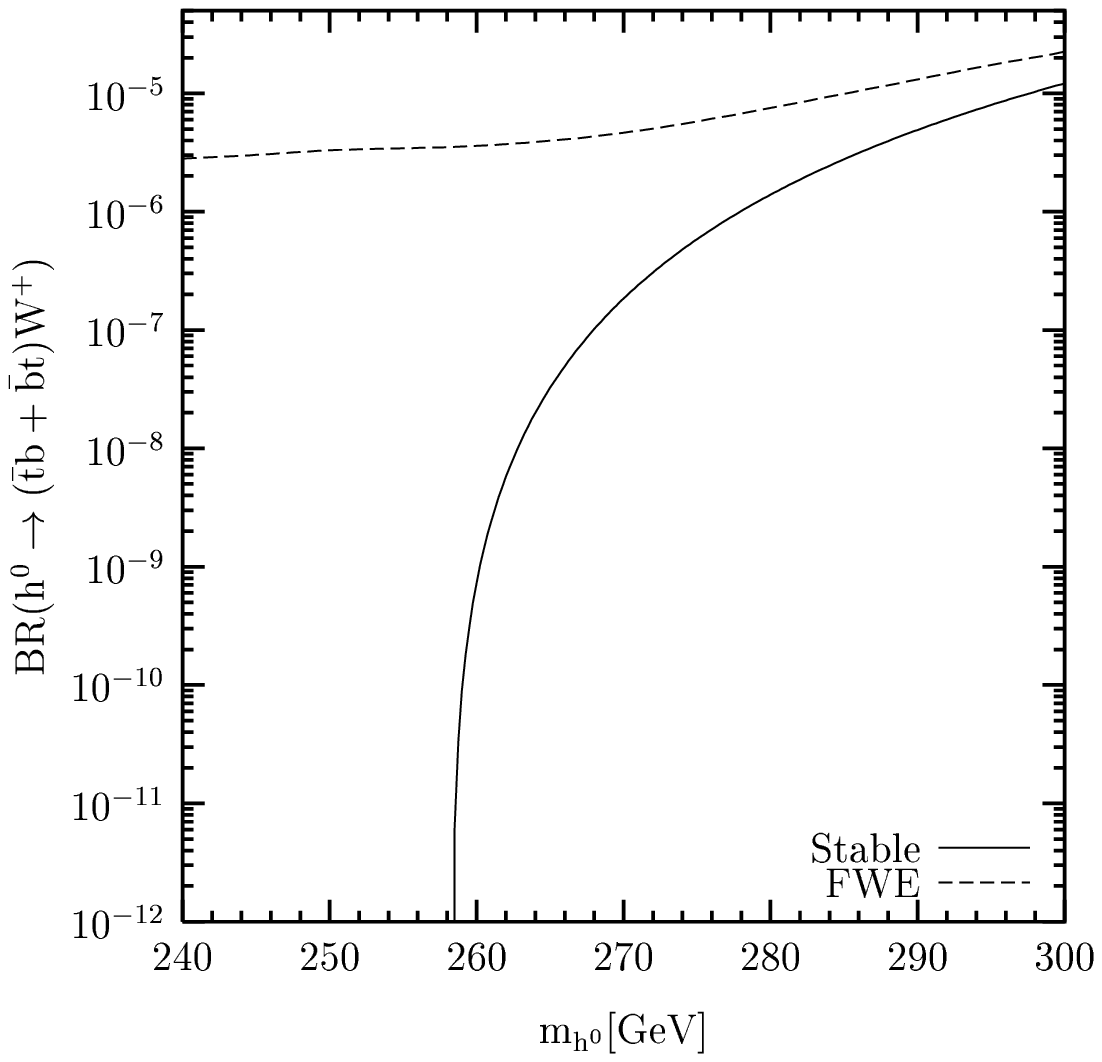 }} }
\vskip -2.3in
    \caption{\emph {The branching ratio
${\rm BR}(h^0 \to t \bar b W^- + \bar t b W^+)$ in model III,
as a function of the $h^0$ mass, 
with FWE (upper curve) and without FWE (lower curve), for $m_t=173.5$ GeV, $m_{H^+}=350$ GeV 
and $\alpha=\pi/4$.}}\label{BRh0}
\end{figure}

\begin{itemize}
 
\item The propagator of the intermediate $W$ is taken from Eq.~(\ref{Vectorprop}).

\item  In the calculation of $\Gamma^0$ in Eq.~(\ref{tV}), the usual sum over the spins of the outgoing top-quark 
is modified to $\sum{u(p_t)\bar{u}(p_t)}=\ptslash+\sqrt{p_{t}^2}$ since,
using the prescription of the CM, the  
final state top-quark is allowed to be off-shell.

\item Throughout the following we assume that the Higgs mass 
spectrum respects 
$m_{h^0} < m_{A^0} \ll m_H^+,~m_H^0$, setting 
$m_H^+=m_H^0=1$ TeV. Thus, the contribution from the charged Higgs exchange, 
i.e., diagram (b) in 
Fig.~\ref{Feynm}, becomes negligible.

\item The total width of $A^0$ is estimated from the decays
$A^0\to \tau\bar{\tau},\,b\bar{b},\,h^0Z,h^0Z^*,\,(t\bar{b}+\bar{t}b)W$, 
and the total width of $h^0$ is estimated from the decays
$h^0\to \tau\bar{\tau},\,b\bar{b},\,W^+W^-,ZZ$. 

\item We set all the relevant flavor diagonal $\lambda$'s 
of the Higgs Yukawa couplings in Eq.~(\ref{yukawa}) to unity, i.e., 
$\lambda_{qq}=1$.

\end{itemize}

With the above assumptions,
the remaining relevant input parameters (in Model III) for evaluating the 
branching ratios under consideration are $m_{A^0}$, $m_{h^0}$ 
and the Higgs mixing angle $\alpha$. 

In Fig. \ref{BRA0} we depict the branching ratio of 
$A^0\to(\bar{t}b+t\bar{b})W$ as a function of $m_{A^0}$, 
for two values of the light Higgs mass $m_{h^0}=170$ and 230 GeV and 
for $m_H^+=1$ TeV, $m_t=173.5$ GeV and $\alpha=\pi/4$. 
We see that near threshold, i.e., $m_{A^0} \sim 260$ GeV, there 
is an enhancement of several orders of magnitude  
due to FWE, wherein the the branching ratio can reach 
${\rm BR}(A^0\to(\bar{t}b+t\bar{b})W) \sim 10^{-2}$. 
Away from threshold, the decay $A^0\to(\bar{t}b+t\bar{b})W$ 
is sensitive to the lightest neutral Higgs mass, $m_{h^0}$.
In this case, 
the inclusion of FWE can increase the branching 
ratio by almost an order 
of magnitude, giving e.g.   
${\rm BR}(A^0\to(\bar{t}b+t\bar{b})W) \sim {\rm few} \times 10^{-1}$ for 
$m_{A^0} \sim 300$ GeV and $m_{h^0}=230$ GeV.    
Thus, FWE in the three-body decay $A^0\to(\bar{t}b+t\bar{b})W$ can 
become very significant -- bringing its BR to the level of tens of percents 
and making it competitive with the $A^0$ two-body decays and, therefore, 
a viable experimental signature for studies of the properties of the 
Higgs sector.

Finally, let us consider the decay $h^0\to(\bar{t}b+t\bar{b})W$. 
In Fig.~\ref{BRh0} we plot its branching ratio as a function of 
$m_{h^0}$ for the same input parameters (of Model III) as in 
Fig.~\ref{BRA0}. In this case, in spite of the large enhancement near 
threshold due to FWE, the ${\rm BR}(h^0\to(\bar{t}b+t\bar{b})W)$ remains 
rather small, i.e., at most of ${\mathcal O}(10^{-5})$, mainly due to 
the much larger $h^0$ total width caused by its tree-level 
decays to a pair of gauge-bosons $h^0 \to WW,ZZ$.    

\section{Summary}\label{}

We have studied and emphasized the importance of FWE 
(finite width effects) 
in decays occurring just around their kinematical thresholds. 
For the inclusion of FWE we have adapted the so called CM
(convolution method). In the CM, the unstable particle 
with 4-momentum $p$ is treated as a 
real physical particle with an invariant mass $\sqrt{p^2}$ and
effectively weighted by a Breit-Wigner-like density function, which,
becomes a Dirac-delta function in the limit that the particle's total 
width approaches zero.   

We first examined the FWE within the SM in the rare and flavor-changing 
tree-level top decay $t \to c W^+W^-$ and then extended our analysis 
to FWE in the tree-level top decays $t \to c W^+W^-$, $t \to c ZZ$ 
and Higgs decays $A^0,~h^0 \to t \bar b W$ in a general two Higgs 
doublets model, the so called Model III, which gives rise to 
tree-level FCNC in the Higgs-fermion sector. 
In all these case we find that FWE can become substantial -- enhancing the
branching ratios for the above decays by several orders of magnitudes 
near threshold. 

Unfortunately, in the SM case, the top decay $t \to c W^+W^-$
remains too small to be of any value in the upcoming high energy colliders, 
i.e., ${\rm BR^{SM}}(t \to c W^+W^-) \sim 10^{-13}-10^{-12}$, 
in spite of the large enhancement due to FWE. 
On the other hand, in Model III, the large enhancement due to FWE in 
all these three-body top and Higgs decays
can make a difference with respect to experimental studies in the upcoming 
hadron colliders.  
In particular, the branching ratios for the top-decays 
$t \to c W^+W^-$ and $t\to c ZZ$ can reach the level of $10^{-4}-10^{-3}$
near threshold -- many orders of magnitudes larger than the corresponding 
branching ratio for the stable W and Z-bosons case (i.e., without FWE).
For the $t\to c ZZ$ decay, the inclusion of FWE is essential since 
such a large branching ratio arises even though the naive threshold 
for this decay is a few GeV away from the most recent 
$1 \sigma$ upper limit on the top mass, $m_t(1\sigma) \sim 180$ GeV.  
 
In the Higgs decays, FWE are more noticeable in the      
pseudo-scalar Higgs decay $A^0 \to(\bar{t}b+t\bar{b})W$, elevating 
its branching ratio to the level of tens of percents, thus 
making this three-body decay channel dominant 
and competitive with its two-body decays 
and, therefore, extremely important for experimental studies.   

Thus, our study shows that FWE is essential 
for a proper treatment of otherwise neglected finite widths of particles 
which emerge at the final state of decays or scattering processes occurring 
just around the threshold.

\begin{acknowledgments}
I.T. would like to thank the HEP group members  at Technion 
for their support and kind hospitality during his stay there. 
The work of G.E. was supported in part by the United States Department 
of Energy under Grant Contract No. DE-FG02-95ER40896 and by 
the Israel Science Foundation. The work of M.F. is supported in part by 
NSERC under grant number 0105354. I.T. also thanks Marc Sher for useful conversations. 
\end{acknowledgments}


\end{document}